**Investigating Drivers' Head and Glance Correspondence**


Joonbum Lee [a], Mauricio Muñoz [a,b,c], Lex Fridman [a], Trent Victor [d], Bryan Reimer [a], and Bruce Mehler [a]

joonbum@mit.edu; ammd@mit.edu; fridman@mit.edu; trent.victor@volvocars.com; reimer@mit.edu; bmehler@mit.edu

[a] MIT AgeLab and New England University Transportation Center, 77 Massachusetts Avenue, Cambridge, MA 02139, USA

[b] TU Munich, LMU Munich, Arcisstraße 21 D-80333, Munich, Germany

[c] University of Augsburg, Universitätsstraße 2, 86159 Augsburg, Germany

[d] SAFER Vehicle and Traffic Safety Centre at Chalmers, SE-402 78, Göteborg Sweden

Corresponding Author:

Joonbum Lee

1 Amherst St., E40-215

Cambridge, MA 02139

Email: joonbum@mit.edu

Phone: 617-715-2011




**ABSTRACT**

The relationship between a driver's glance pattern and corresponding head rotation is highly complex due to its nonlinear dependence on the individual, task, and driving context. This study explores the ability of head pose to serve as an estimator for driver gaze by connecting head rotation data with manually coded gaze region data using both a statistical analysis approach and a predictive (i.e., machine learning) approach. For the latter, classification accuracy increased as visual angles between two glance locations increased. In other words, the greater the shift in gaze, the higher the accuracy of classification. This is an intuitive but important concept that we make explicit through our analysis. The highest accuracy achieved was 83% using the method of Hidden Markov Models (HMM) for the binary gaze classification problem of (1) the forward roadway versus (2) the center stack. Results suggest that although there are individual differences in head-glance correspondence while driving, classifier models based on head-rotation data may be robust to these differences and therefore can serve as reasonable estimators for glance location. The results suggest that driver head pose can be used as a surrogate for eye gaze in several key conditions including the identification of high-eccentricity glances. Inexpensive driver head pose tracking may be a key element in detection systems developed to mitigate driver distraction and inattention.





**Highlights**

- After applying four machine learning algorithms to classify two glance locations (forward vs. center stack), the Hidden Markov Model provided the best accuracy at 83%.

- Random Forest model reached accuracy at 90% to classify glances to forward from glances to the right mirror.

- Increasing visual angles between two glance locations may help to increase classification accuracy.

**ACKNOWLEDGEMENTS**

Support for this work was provided by the US DOT's Region I New England University Transportation Center at MIT, The Santos Family Foundation and the Toyota Class Action Settlement Safety Research and Education Program. The views and conclusions being expressed are those of the authors, and have not been sponsored, approved, or endorsed by Toyota or plaintiffs' class counsel. An earlier version of this work (Muñoz, Lee, Reimer, Mehler, & Victor, 2015) appeared in the proceedings of the 8th International Driving Symposium on Human Factors in Driver Assessment, Training, and Vehicle Design. Also authors would like to acknowledge the contribution of Shannon Roberts who provided valuable comments on this manuscript.



**1. Introduction**

Eye movements have long been studied in the context of driver behavior, attention management, and task related visual demand assessment (e.g., Wierwille, 1993). Although eye tracking systems have been employed in numerous scientific studies (e.g., Wang, Reimer, Dobres, & Mehler, 2014), the technology is susceptible to data quality issues (Ahlstrom & Victor, 2012; Sodhi, Reimer, & Llamazares, 2002) and difficult to reliably use in production systems. Research on the correspondence between eye and head movement suggests that head pose data may be useful as a surrogate for eye-glance data (e.g., Talamonti, Huang, Tijerina, & Kochhar, 2013; Talamonti, Kochhar, & Tijerina, 2014). Talamonti and his colleagues (2013) found a low likelihood (65% or less) of head turns when glancing to the instrument panel and rearview mirror, and high likelihood (93% or more) when glancing to the left mirror, center console, and center stack. Talamonti (2014) suggested that driver-specific thresholds need to be set in order to meaningfully use head yaw data as a glance predictor. The previous studies utilized a fixed-base driving simulator to collect data and applied a simple classifier to understand relationship between head turns and glance locations.

The present study aims to further investigate whether head-rotation data can be used as a surrogate for eye-glance behaviors in on-road data. Head rotation and glance data were drawn from a study conducted by the Virginia Tech Transportation Institute (Transportation Research Board of the National Academies of Science, 2013). This study utilized the data to: (1) begin developing a deeper understand of how drivers rotate their heads, (2) investigate individual differences in head-glance correspondence, and (3) generate input features for classifiers that predicted glance allocations. Based



upon the literature noted above, it was expected that head-rotation data could be used to predict some, but not all, glances away from the road. Therefore, as an initial effort, and approaching the problem from a classification perspective, we tested whether head rotation can predict glances to the forward road, to the vehicle's center stack (e.g., climate controls, infotainment display), and to other key locations in the vehicle (e.g., instrument cluster, mirrors, center stack, etc.). Subsequent efforts then evaluated the degree to which machine learning algorithm could predict glances to other closer and farther regions of the vehicle interface and to evaluate the degree to which individual differences influence behavior.

The present study has two main objectives: (1) to investigate the use of head pose data to predict glance location; (2) to understand the potential individual differences in head rotation patterns. To achieve these objectives, we built a framework utilizing principal component analysis (PCA) and machine learning techniques by considering several factors that may affect model performance and interpretation.

## 2. Methods

This study is a secondary analysis of a subset of data collected by the Virginia Tech Transportation Institute (VTTI) in support of the Strategic Highway Research Program 2 (SHRP 2) naturalistic driving study (Transportation Research Board of the National Academies of Science, 2013). The data were provided to the MIT AgeLab under an IRB approved data sharing agreement. A total of 44 participants were available (22 participants for static trials and 22 participants for dynamic trials). The sample spans four age groups (18-35, 36-50, 51-65, and over 66, with a majority of



cases falling in the first two groups). Participants were initially recruited to ensure that the data set represented a wide array of facial geometry. Approximately twice as many males ($N$ = 30) than females ($N$ = 14) were tested. Participants who met the study's eligibility criteria were assigned to participate in either static trials (e.g., data collected while not driving) or dynamic trials (e.g., data collected while driving). Data were collected in a 2001 Saab 9-3 instrumented with a data acquisition system to collect a number of metrics, including digital video of the drivers face. This video was recorded by a camera mounted below the rearview mirror. The present study focused on analysis of the dynamic trials as an earlier report (Muñoz et al., 2015) showed limited overlap between the distribution of head rotations associated with glances to the road and center cluster in those trials. Thus, in static conditions glances between the road and center cluster appear more easily separable based upon head position than during dynamic trials. As such, 22 participants from the dynamic trials were analyzed and included in this analysis.

*2.1. Test Trials*

The dynamic trials were conducted on a predefined route around Blacksburg, Virginia. This route was approximately 15 miles in length and consisted of various road types (e.g., two lane road, residential, rural, and divided highway). During the session, participants were instructed to perform a set of five basic tasks: (1) report current vehicle speed, (2) indicate if any vehicles are immediately adjacent to the test vehicle, (3) turn the radio on and then off, (4) locate the cell phone in the center console and (5) complete a brief simulated cell phone conversation.



*2.2. Data Reduction*

Video of each task/glance was recorded at 15 frames per second and decomposed into frames for analysis. Each video frame was annotated by two independent analysts who labeled seven predefined facial landmarks: (1) outer corner of the participant's right eye, (2) inner corner of the participant's right eye, (3) outer corner of the participant's left eye, (4) inner corner of the participant's left eye, (5) the tip of the participant's nose, (6) the right corner of the participant's mouth, and (7) the left corner of the participant's mouth. Two analysts' x and y pixel coordinates for each landmark were averaged, and if the average frame pixel correction exceeded 3.5 pixels, the frame was considered as a significant disagreement between two analysts, and was excluded from the rotation estimate data set. If either analyst could not make a reliable annotation, the landmark was marked as "missing", and the frame was excluded from the rotation estimate data set. For each video frame, geometric methods (e.g., Murphy-Chutorian & Trivedi, 2009) which utilize feature locations (e.g., eyes, mouth, and nose tip) were used for head rotation estimation. The head pose data consisted of three rotation estimates (i.e., X, Y, and Z rotation). Figure 1 shows a rotation coordinate system.



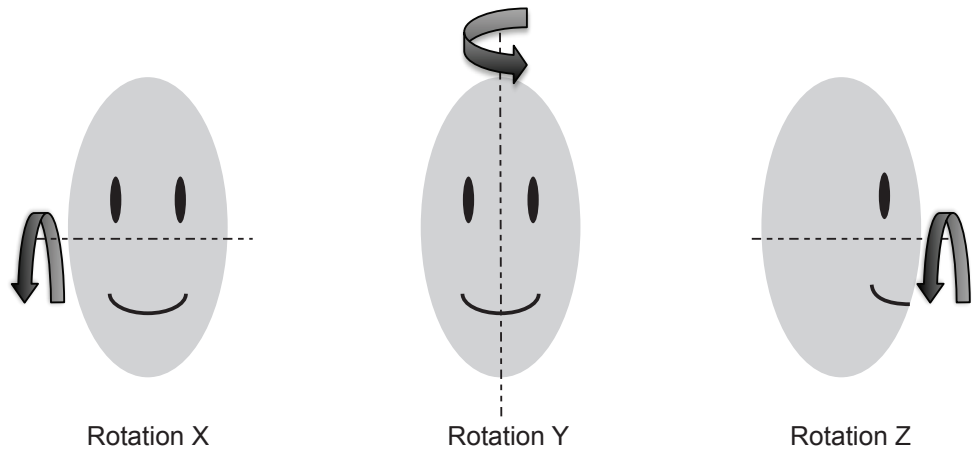

**Figure 1. Head rotation coordinate system.**

Glance locations were coded by trained video analysts on a frame-by-frame basis into one of 16 locations: forward, left forward, right forward, rearview mirror, left window/mirror, right window/mirror, over-the-shoulder, instrument cluster, center stack, cell phone, interior object, passenger, no eyes visible—glance location unknown, no eyes visible—eyes are off-road, eyes closed, and other. Figure 2 shows 10 of the 16 glance locations. A senior analyst reviewed the output of the coding and provided feedback to less-experienced analyst. Glance allocations for each subject and task were merged with head rotation data using timestamps.



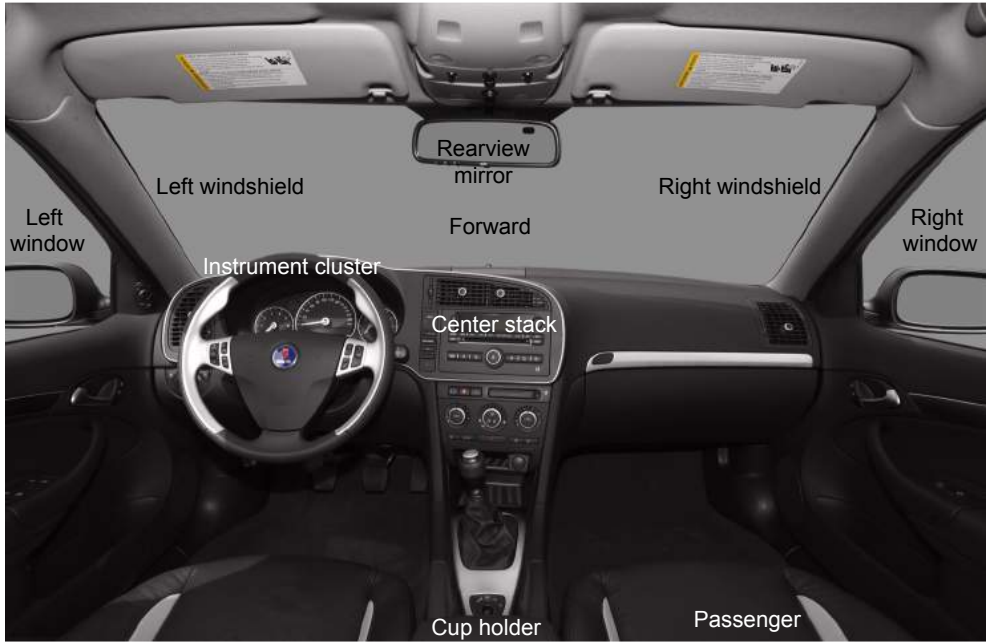

**Figure 2. Glance locations for the manual coding.**

*2.3. Model Training and Validation*

Training data were derived from the data set by taking all data belonging to a randomly sampled subset of all available subjects (80%). The remaining (20%) subjects were used to build a validation data set. As one of the tested classifiers, i.e., the Hidden Markov Model (HMM), takes the temporal structure of the input data into account, the timestamp ordering of the samples for each subject were maintained.

All rotation variables were individually normalized by computing their individual z-scores across all subjects. The performance measures reported in the result section were computed using this normalization method. Furthermore, to decrease potential variances from subject sampling, a Monte-Carlo sampling technique was used. For each of 50 iterations, train and test sets were generated as above. All models were trained, and performance values were then computed for each classifier as the mean of



each performance metric (standard accuracy, F1 score, and Kappa value) over all iterations.

One key issue considered is the unbalanced class structure (i.e., skewness) of the dataset, as glances to the forward roadway heavily outnumber glances to any other single location within the vehicle. For instance, out of all the glances to the forward roadway and center stack, approximately 95% belong to the former class. Subsampling was used to prune away the over-represented glance locations in the data.

*2.4. Data Exploration*

The intrinsic discriminative quality of the data plays a crucial role in any classification framework, i.e., classification may be difficult in datasets in which classes overlap strongly in feature space. To explore this aspect of our data, visual representations of the most salient patterns were developed using PCA. PCA is a common statistical technique for pattern detection, data visualization, and compression (Jolliffe, 2005). It is used here to represent the raw data in terms of its salient structural patterns, computing first the covariance matrix across all variables, and then extracting the eigenvectors of this matrix. Given this information, which characterizes how statistical variance is distributed amongst combinations of variables, this analysis can identify and visualize properties that might have an impact on classification performance, in particular which variables are most likely to contribute to the classification procedure.



*2.5. Model Development*

The classification methods presented in this paper are (1) k-Nearest Neighbor, (2) Random Forest, (3) Multilayer Perceptron, and (4) Hidden Markov Models. The parameters of each model were tuned with an experimental validation set (i.e., a random subset of the larger data pool). These methods were chosen for based on the trade-off in running time, space complexity, and difficulty of parameter tuning. k-Nearest Neighbor (kNN) algorithm has the lowest number of parameters but the highest space and running time complexity requirements during evaluation, but is very fast during training. Random Forest classifier (Breiman, 2001) is a representative ensemble method with high space complexity requirements both for training and evaluation, but unlike kNN it is both fast to train and fast to evaluate. Random forest uses a random subset of each input sample at different nodes to train the corresponding weak learner. This has the added benefit that as training progresses, variables with low information content are automatically filtered out, thus making the classifier especially well-suited for data structured across heterogeneous input variables. Multilayer Perceptron (MLP) was taken as a representative of the larger class of Artificial Neural Networks (ANN) for their ability to model non-linear relationships between data points. MLP space complexity is low for both training and evaluation, while running time is slow for training and fast for evaluation. Hidden Markov Models (HMMs) (Rabiner & Juang, 1986) are employed to test how much of the classification signal lies in the temporal structure of the data. Sequences of head rotation and glance duration features are fed to the classifier, which then infers a single class label for the sequence of samples. As in the classical



approach, one HMM is built from data from each class (glance location). The class label of an unobserved sequence is then determined by finding the HMM and its corresponding class that maximizes the log probability of the test sequence.

*2.6 Model Performance Measures*

To assess performances of the classifiers, three performance measures were used in this study:

1. Classification accuracy (AC) (Sokolova & Lapalme, 2009): the percentage of correctly classified samples (or sample sequences for the HMM classifier):

$$Classification\ Accuracy = \frac{Number\ of\ correctly\ labeled\ samples}{Total\ number\ of\ samples}$$

2. F1-score (FS) (Sokolova & Lapalme, 2009): a measure of how well the classifier was able to distinguish between classes given an unbalanced dataset.

$$F1\ score = \frac{2 \times (Positive\ predictive\ value \times Sensitivity)}{Positive\ predictive\ value + Sensitivity}$$

$$Positive\ predictive\ value = \frac{Number\ of\ true\ positives}{Number\ of\ positive\ calls}$$

$$Sensitivity = \frac{Number\ of\ true\ positives}{Number\ of\ true\ positives + Number\ of\ positive\ calls}$$

3. Cohen's Kappa statistic (KP) (Carletta, 1996): a measure indicating how well a classifier agrees with a perfect predictor (higher values indicate high agreement).

$$Cohen's\ kappa = \frac{P(A) - P(E)}{1 - P(E)}$$



$P(A) = \textit{Relative observed agreement between model and perfect predictor}$

$P(E) = \textit{Probablity of chance agreement between model and perfect predictor}$

## 3. Results

To answer the key questions outlined: (1) PCA was applied to the driving data as a method of quantifying the contributions of each head angle (X, Y, and Z) in their ability to discriminate between glance locations (e.g., forward vs. center stack), (2) several predictive models were tested for predicting glance location based upon head position while driving and their accuracies compared, and (3) individual differences in head-glance correspondence during driving were addressed.

### 3.1. Principal Components Analysis

PCA was used to reinterpret the X, Y, and Z filtered rotation variables along the salient variance properties of the data. Figure 3 (a) plots the dynamic data (center stack vs. forward) in terms of two principal components (principal component 1 and 2). Each axis of the graph corresponds to each principal component and represents salient statistical behavior of the data along that component. Results in Figure 3 (b) and Figure 4 (b) showed that the third component was responsible for only a trace amount of the total variance in the data. Therefore, Figure 3 (a) and Figure 4 (a) show the data along only the first two components of the PCA decomposition. The distribution of individual data points and their class correspondence in Figure 3 (a) was compared with the actual principal component values in Figure 3 (b) to establish an informal overview of which variables are most likely to contribute to the classification effort. For instance, a



rough clustering of forward glances may be observed in Figure 3 (a). This cluster center

lies at moderate to high values of Principal Component 1 (PC1) and slightly negative

values of Principal Component 2 (PC2). In Figure 3 (b), the X rotation variable most

adequately fits this profile, suggesting that X rotation holds the strongest classification

signal for the center stack vs. forward roadway classification problem. Figure 3 (b)

likewise shows that X rotation, as the variable with highest absolute presence in the first

principal component (column in the matrix), has the largest statistical variance (loosely,

information content) from amongst all three variables, which further supports this idea.

The same analysis was made for the center stack vs. right mirror case. Figure 4

provides a 2-dimensional sample distribution plot as well as the corresponding principal

components. It may be observed that most right mirror samples cluster around high

values of PC1 and negative values of PC2. In contrast to the previous case, this profile

correlates best with the Y rotation variable (horizontal head movement), as would be

expected, which in this case also has the highest variance/information content from

amongst all three variables. As such, Y rotation is likely to be the most significant

variable in the center stack vs. right mirror classification problem.

     The fact that there is no clear clustering of the data suggests that the

classification problem is difficult and mostly dependent on a single variable only for

cases of a significant increase in the visual angle. Rather, a mix of X and Y rotation

variables were identified by the PCA analysis to be the primary contributors to the

classification according to the amount of explained variance.



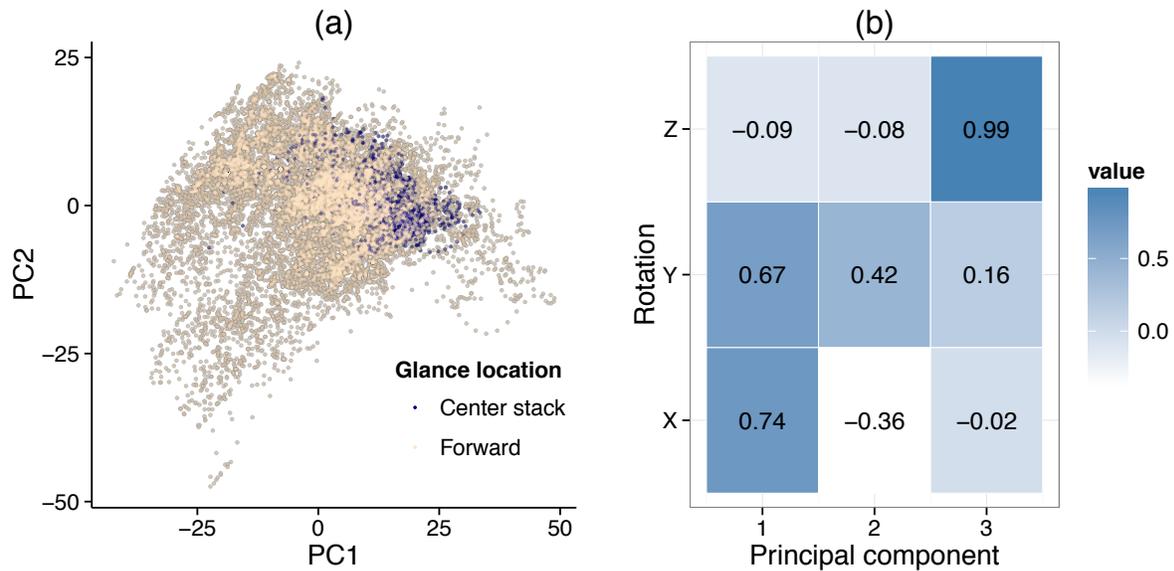

**Figure 3. Principal component analysis (PCA) of dynamic data, using head rotation: (a) Glances to the center stack and forward roadway. (b) Principal components of head rotation X, Y, and Z [The values have been averaged over a total of 50 iterations of randomly generated training sets (80% of all subjects were randomly sampled for each set)].**

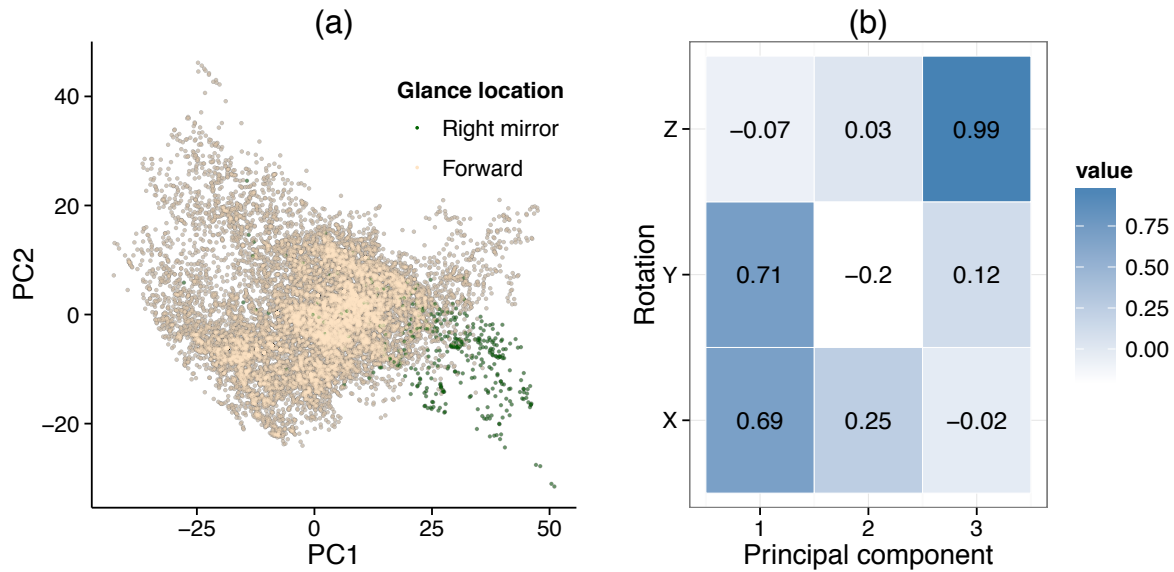

**Figure 4. Principal component analysis (PCA) of dynamic data, using head rotation: (a) Glances to the right mirror and forward roadway. (b) Principal components of head rotation x, y, and z [The values have been averaged over a total of 50 iterations of randomly generated training sets (80% of all subjects were randomly sampled for each set)].**



*3.2. Model Validation*

Table 1 presents the performance measures for all four classifiers using the two representations of the data (balanced vs. unbalanced) for the forward vs. center stack case. As noted earlier, Monte-Carlo sampling (50 iterations) was applied. Using the balanced dataset that removed glance distribution bias during training leads to a higher performance in terms of sensitivity/specificity (F1 scores all ≥ 0.68) and prediction quality (Kappa all ≥ 0.41) of each classifier compared to (F1 scores all ≥ 0.04) and (Kappa all ≥ -0.07) for the original unbalanced data. The HMM classifies sample sequences corresponding to blocks of data within a subject and task group. The relatively strong performance of the model suggests that the temporal structure of head rotation features is another potential source of information. Overall, there is a general consensus amongst all classifiers regarding the discriminative quality of head rotation data. Though all classifiers using the balanced dataset perform better than a chance predictor, there is a clear upper bound on how much these features contribute to classification.

**Table 1. Performance measures (AC: accuracy, FS: F1 score, KP: Kappa statistic) across all classifiers and class distributions for dynamic data, forward roadway vs. center stack**

|  | Original Dataset | | | Balanced Dataset | | |
|---|---|---|---|---|---|---|
|  | AC | FS | KP | AC | FS | KP |
| **k-Nearest Neighbor** | 0.93 | 0.19 | 0.16 | 0.80 | 0.80 | 0.59 |
| **Random Forest** | 0.86 | 0.30 | 0.24 | 0.79 | 0.78 | 0.59 |
| **Multilayer Perceptron** | 0.78 | 0.29 | 0.22 | 0.80 | 0.82 | 0.59 |
| **Hidden Markov Model** | 0.84 | 0.28 | 0.22 | 0.83 | 0.68 | 0.57 |



In addition, other locations within the vehicle were also tested against glances to the forward roadway in order to examine the relationship between classification accuracy and the visual angle of the target. HMM and Random Forest models, which showed relatively higher accuracy among other classifiers, were selected and tested. Table 2 places the previous center stack classification efforts in this context and gives performance measures for the two classifiers with the overall best performance. As expected, a rough correlation between increasing visual angle and classification accuracy may be observed, reaching up to 90% classification rate with a balanced data set. The results may support that head pose data can detect particularly detrimental glances (i.e., high-eccentricity glances) with high accuracy, whereas using head pose data alone does not provide high accuracy to detect low-eccentricity glances.

**Table 2. Performance measures (AC: accuracy, FS: F1 score, KP: Kappa statistic) across class distributions for the Random Forest and HMM classifiers for dynamic data, forward roadway vs. instrument cluster, vs. left mirror, vs. center stack, vs. right mirror**

| Forward vs. | Model | Original Dataset | | | Balanced Dataset | | |
|---|---|---|---|---|---|---|---|
| | | AC | FS | KP | AC | FS | KP |
| Instrument Cluster | Random Forest | 0.59 | 0.33 | 0.08 | 0.56 | 0.48 | 0.11 |
| Instrument Cluster | Hidden Markov Model | 0.66 | 0.32 | 0.12 | 0.66 | 0.61 | 0.33 |
| Left Mirror | Random Forest | 0.86 | 0.30 | 0.24 | 0.79 | 0.78 | 0.59 |
| Left Mirror | Hidden Markov Model | 0.84 | 0.28 | 0.22 | 0.83 | 0.68 | 0.33 |
| Center Stack | Random Forest | 0.85 | 0.77 | 0.66 | 0.83 | 0.81 | 0.66 |
| Center Stack | Hidden Markov Model | 0.83 | 0.72 | 0.60 | 0.85 | 0.83 | 0.69 |
| Right Mirror | Random Forest | 0.95 | 0.74 | 0.72 | 0.90 | 0.89 | 0.80 |
| Right Mirror | Hidden Markov Model | 0.93 | 0.69 | 0.65 | 0.87 | 0.73 | 0.66 |



*3.3. Individual Differences in Head-Glance Correspondence*

Also individual differences in head-glance correspondence were tested. To minimize potential variability from characteristics of tasks, only the radio task (e.g., "Turn the radio on and then off"), which required glances to the center stack from the dynamic setting, was selected and analyzed. Figure 5 illustrates the distribution of 21 participants' individual Y rotation while glancing to the center stack during the radio tasks (there was one subject who did not glance to the center stack and that case was excluded for the subsequent analysis). As can be observed in Figure 5, a wide range of Y rotations exists while glancing to the center stack across the subject pool, with some subjects showing relatively narrow distributions and others showing wide distributions. It is also important to observe that the center point of each subject's distribution varies even they are looking at the same object in space.



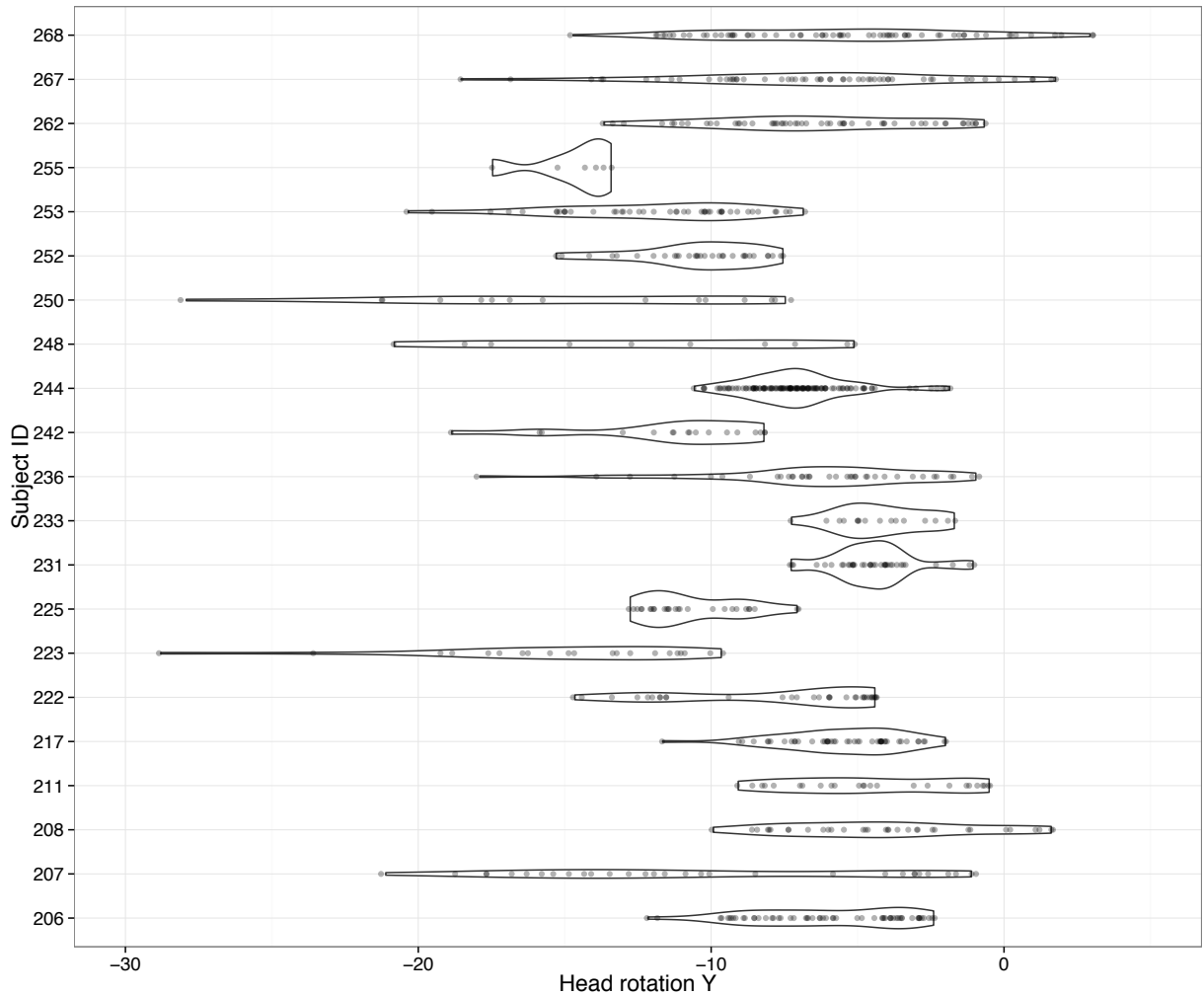

**Figure 5: Comparison of individual distribution of Y rotation while glancing to the center stack during the radio task.**

To further explore the differences in rotation distributions in glances to the center stack in relation to glances to the forward roadway, Y rotations were plotted over time while completing the radio task (see Figure 6) for an illustrative sample of three subjects. This figure visualizes how drivers horizontally rotate (e.g., Y rotation) their head while engaging in the radio task and their glance locations over time (differentiated in colors). The top frame of Figure 6 illustrates a profile that has relatively narrow range of Y rotation while glancing to the center stack, and (relatively) limited overlap between the ranges of Y rotation corresponding to glances to forward and glances to the center



stack. The middle frame of Figure 6 illustrates a profile that covers a wider range of Y

rotation with significant overlap of the two glance locations. Finally, the lower frame

illustrates a profile with a narrow range of Y rotation with a sizable overlap between the

glance locations.

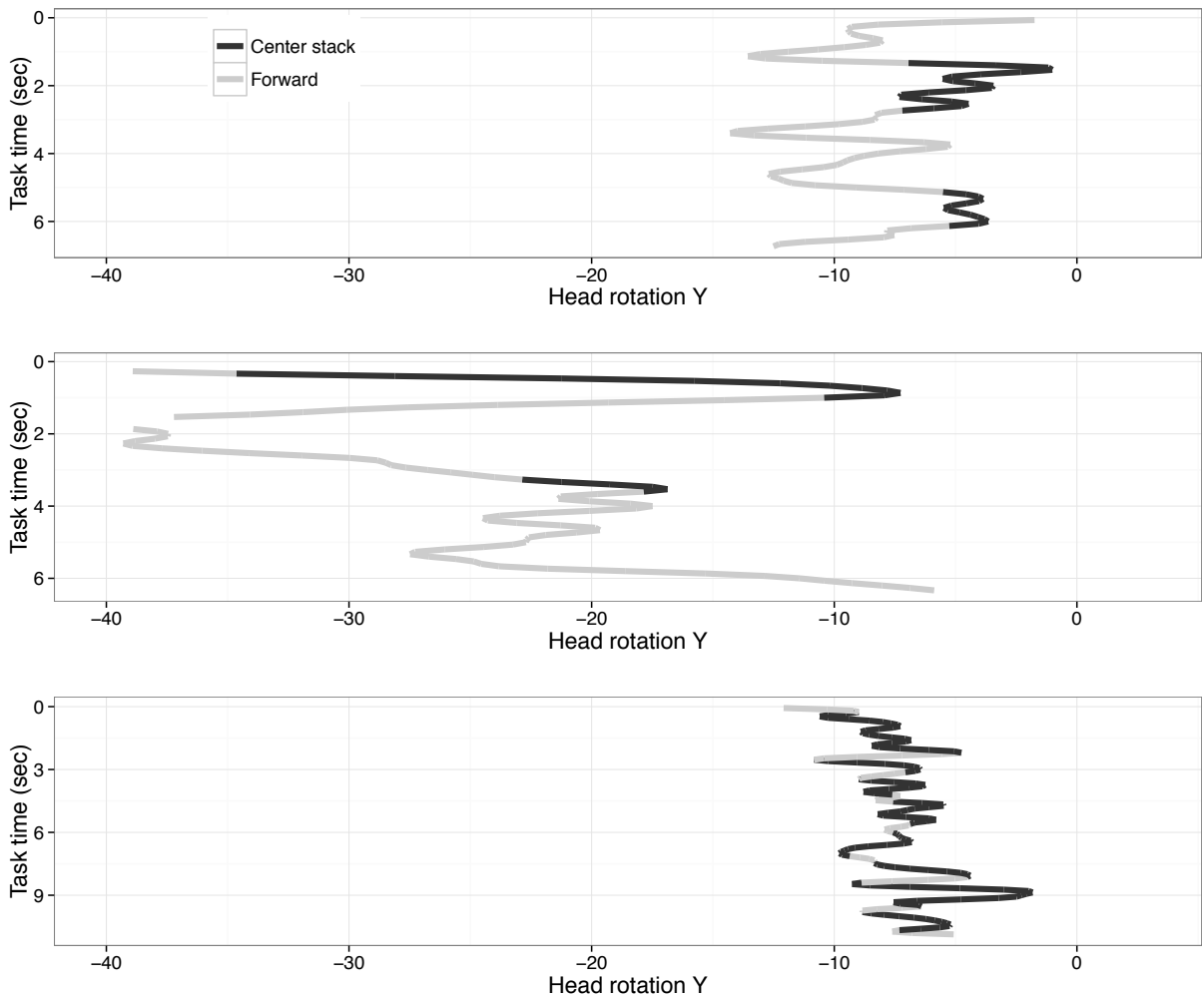

**Figure 6: Illustration of three subjects' Y rotation over time during dynamic radio tasks (note: line color represents glance locations).**

Based on the exploratory findings, it was assumed that individual difference in

head-glance correspondence may exist. Figure 7 shows 21 subjects on two dimensions:

(1) the mean difference of Y rotation between glances to forward and to the center



stack, and (2) the range of Y rotation (i.e., distribution width of rotation Y while glancing to the center stack). The result showed that the two dimensions were positively correlated [$r$ (19) = .73, $p$ < .001], indicating that subjects who showed wider ranges of horizontal head rotations tended to have higher mean differences of rotation Y while glancing to forward and the center stack. For example, subject 244 and 225 showed relatively narrow ranges of horizontal head rotations (less than 10 degrees) while glancing to the center and their mean rotation angles for glancing to the center stack were relatively close to their mean rotation angles for glancing to forward (the mean differences were 1.05 degrees for subject 244 and 2.16 degrees for subject 225). This may indicate that subjects on the left-bottom area in Figure 7 such as subject 244 and 225 (i.e., narrow width and small mean difference) moved their head less actively to glance to the center stack, whereas subjects on the right-top are actively moved their head to glance to the center stack location.

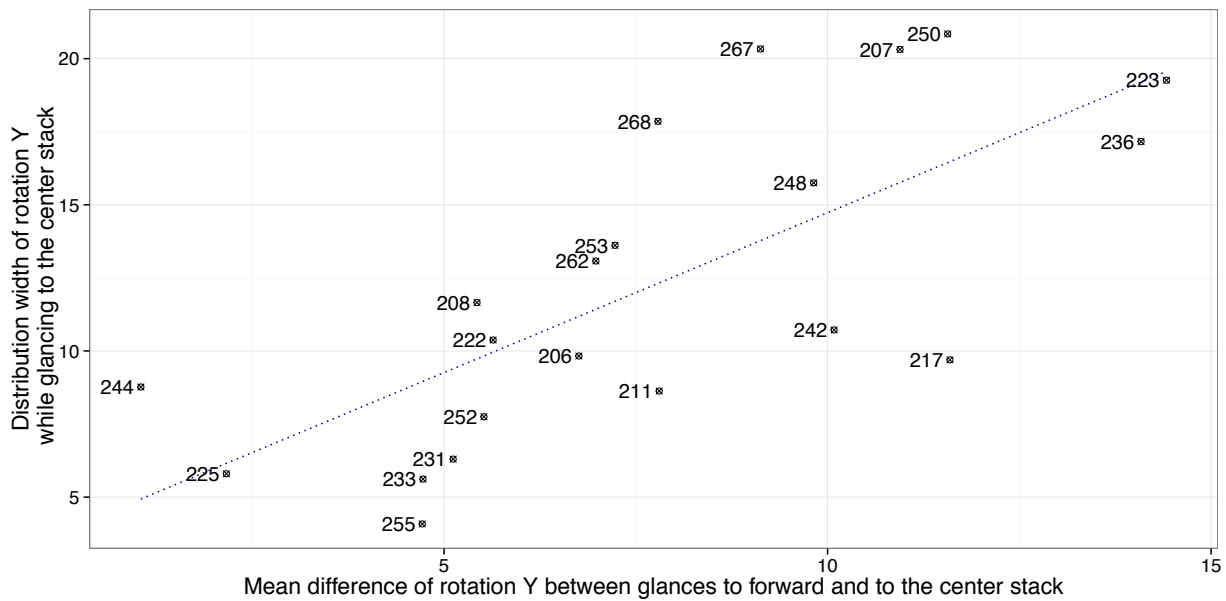

**Figure 7: Drivers' head angle profiles while glancing to the center stack during the radio tasks (note: numbers represent subject ID).**



**4. Discussion**

This study investigates the relationship between head rotation and glance behavior during on-road driving. Various machine learning techniques were employed to examine the predictive value of head rotation for glance location at an individual level. PCA analysis revealed that a combination of X and Y head rotation variables is the primary contributor to classify forward glances from glances to the center stack and right mirror. That is, both vertical and horizontal head rotations are key variables to classify glance locations.

A total of four classifiers from a wide range of data interpretation techniques were used to detect patterns in head rotation data. Both unbalanced raw data, which included more cases of glancing forward than glancing to the center stack, and balanced data were tested. Substantial performance gains were observed when using the balanced training data set. For the forward roadway vs. center stack case, Hidden Markov Models performed the best with an accuracy of 83%. All of the modeling approaches provided results that were well in excess of chance findings, suggesting that head rotation data is a fairly robust predictive signal. Given that the limited number of glances to non-forward locations (i.e., glances to the center stack accounted for less than 5% of the total glances recorded) were captured during short/simple secondary tasks, model performance may be best considered as relative lower bound on the possible predictive quality. Given the time series nature of more complex glance allocation strategies and performance of the Hidden Markov Model, higher predictive accuracies may be achievable. This study also looked at the variability in classification accuracy with



increasing visual angles to show a significant correlation between the accuracy and visual angles.

There may be multiple factors that influence drivers' head-glance correspondence such as: (1) road environment (e.g., highway driving vs. rural driving), (2) secondary-task characteristics (e.g., tasks require long off-road glances from drivers vs. tasks require short off-road glances), (3) individual strategies to interact with secondary tasks (e.g., fixing a head to forward while glancing to the center stack), and (4) physical constraints. For this reason, we analyzed only one type of the secondary tasks (i.e., the radio task) for testing individual differences (note that only this analysis subsampled data for one task and other analyses used an entire data set including all tasks). The result showed that individual differences in head-glance correspondence may exist. It is well known that owls have to turn their entire head to change views as their eyes are fixed in their sockets, whereas some lizards (such as Chameleons) have very large angles of eye movement. We also found lizard type drivers (e.g., subject 244 and 225 in Figure 7) and owl type drivers (e.g., subject 223 and 236 in Figure 7), and it was expected that head pose data could predict glance regions with higher accuracy for the owl type drivers (i.e., active head movers). This result suggests the need for a user-specific model (e.g., training a classifier for each individual to detect glances away from the road by using head rotation) or additional input features (i.e., other facial features or pupil location) to increase model performance, especially for the lizard type drivers (who barely move their head while glancing away from the road). Furthermore, efforts should assess the predictive power of head rotation data for certain types of glances such as those that are of longer duration and linked to greater risk of collision (Victor et al., 2015)



**5. Conclusion**

The present study investigated head pose data to test the feasibility of using head pose to predict glance location. This study also systematically tested factors that may affect model performance (e.g., data structure, visual angles between two glance locations, and individual differences in head-glance correspondence). This study achieved fairly accurate classification performance (e.g., classifying glances to forward vs. glances to the center stack), and supports the feasibility of detecting drivers' glances away from the road by not using eye-tracking data. Especially, head pose data accurately classified glances to farther regions (i.e., high-eccentricity glances) from the center forward region. The work suggests that individual differences in head-glance correspondence may be separate into two classes. However, from the data that is available, it is not clear if an individual can be "assigned" to one of the two classes, i.e. they are an "owl" or "lizard", or if there are more factors such as roadway conditions, secondary type interacting with some individual propensity for certain movement patterns.

This study used manually coded on-road data, which are relatively more valid and reliable than automatically tracked eye/head data from a driving simulator. Overall, this work suggests that head rotation data, a feature that may be recorded in the vehicle with limited sophistication using commercially available sensors, may provide a potentially lower cost and higher quality estimate of attention allocation than eye tracking data. Head movements may be used to fairly reliably predict safety critical off-road glances to regions in the vehicle frequently associated with in-vehicle distractions.